# 2D-Material-Assisted Bistable Switching of Gap Plasmons Disclosed By Femtosecond Pulse Scattering Spectra

Tian Yang[1]*, Hui Yi[1], Xiaodan Wang[1], Yichen Miao[1], Cheng Chen[1], Jing Long[1], Xiangyang Kong[2], Lin Wu[3]

[1]State Key Laboratory of Advanced Optical Communication Systems and Networks, Key Laboratory for Thin Film and Microfabrication of the Ministry of Education, Department of Instrument Science & Engineering, School of Electronic Information and Electrical Engineering, Shanghai Jiao Tong University, Shanghai 200240, China.

[2]School of Material Science and Engineering, Shanghai Jiao Tong University, Shanghai 200240, China.

[3]Institute of High Performance Computing, Agency for Science, Technology, and Research (A*STAR), 1 Fusionopolis Way, #16-16 Connexis, Singapore 138632.

**Abstract:** Nanosphere-on-mirror plasmonic antennas, each having a monolayer graphene or $MoS_2$ sheet in the gap, were pumped with a femtosecond laser. Abrupt turnings in the scattering linewidth and peak intensity trends as the laser power changed were experimentally observed. Theoretical modelling of dynamic plasmon evolvement attributes the turning to transitioning between two plasmon states, with a universal switching threshold of four-wave mixing efficiency $\approx 0.14\%$. This bistability is rendered by a strong feedback from the nonlinear gap current to the gap plasmons, and involvement of both four and high-order wave mixing. This work reveals a pathway to making energy efficient nonlinear plasmonic elements.

Nanoplasmonic devices are promising candidates for building ultracompact and broadband nonlinear optical devices, since they can confine electromagnetic fields to sub-wavelength volumes so that phase matching requirement is removed, and to high intensities so that nonlinear processes are boosted [1]. Four wave mixing (FWM) [2,3], second harmonic generation [4-6], third harmonic generation (THG) [7-9], modulation [10-12], and supercontinuum generation [13-15] have been reported for them, with proposals of applications in nonlinear metamaterials and metasurfaces [16-18], optical parametric amplification [19] and sensing and imaging [20,21]. In particular, since their footprints reach down to that of CMOS transistors, they were envisaged as a promising candidate for integrating nonlinear optical functions into high density integrated circuits [22]. However, the extremely low efficiencies and/or high operating powers of these devices have been a critical obstacle that hinders their applications. For example, a micrometer length wave-guiding device has been reported with efficient FWM (signal-to-idler conversion efficiency ~20%) but under laser powers as high as 30 W [3], while sub-micrometer plasmonic devices have been reported with THG and FWM efficiencies less than 0.01% under similar laser intensities [7,8].

On the other hand, nanosphere-on-mirror (NSoM) plasmonic antenna provides an extremely strong confinement of gap plasmons in a well-controlled sub-nanometer height junction [23-25]. It has been extensively investigated for extraordinary light-matter interactions at the nanoscale, including strong coupling and Rabi splitting [26,27], single-molecule vibration mode imaging [28,29], molecular cavity optomechanics [30], phonon stimulated Raman scattering [31-33] and hot electron driven single-molecule chemistry [34-37].

In this work, through observing time-averaged femtosecond laser scattering off the NSoM resonant antennas, whose gaps were filled with a monolayer graphene or $MoS_2$ sheet, we have disclosed features of bistable switching, which were particularly shown as an abrupt turning of the scattering linewidth and intensity as the laser power increased over a threshold. To understand the switching mechanism, we have developed a theoretical model to describe the dynamic evolvement of gap plasmons, which nicely reproduces the experiment results. The formation of bistability is attributed to a strong Purcell feedback from the nonlinear current of the 2D material to the gap plasmons when the antenna is pumped on resonance, and the cooperation of FWM and higher-order wave mixing, in contrast to previous proposals of plasmonic bistability based upon Kerr effect and off resonance pumping [38-40]. The switching laser power for the $MoS_2$-embedded antenna was as low as ~38 mW, showing a performance leap for plasmonic nonlinear nanodevices in terms of efficient functionality and low energy consumption. In addition, the experiment method and theoretical model developed in this work show a useful approach to identify state transitioning during an ultrafast pulse indirectly, without having to resolve the technical challenge of measuring a hysteresis loop with ultrahigh time resolution.

**Experiments—**A schematic of the plasmonic antenna switching device and the optical experiment is shown in Fig. 1A. Each individual antenna consists of a 100 nm diameter gold nanosphere sitting on top of an atomically flat opaque gold plane. The nanosphere and its mirror image in the gold plane form a vertically oriented plasmonic antenna. Between the nanosphere and the mirror plane is a monolayer 2D material sheet, thus defining a sub-nanometer gap. Such an antenna will produce a gap hotspot where the amplitude of electric field is resonantly enhanced by more than two orders of magnitude compared to that of the incident laser [24,25,41]. A ~5nm linewidth fs laser is focused upon the antenna, and the time-averaged scattering spectra are taken at different laser powers to investigate the features resulting from nonlinear wave mixing.

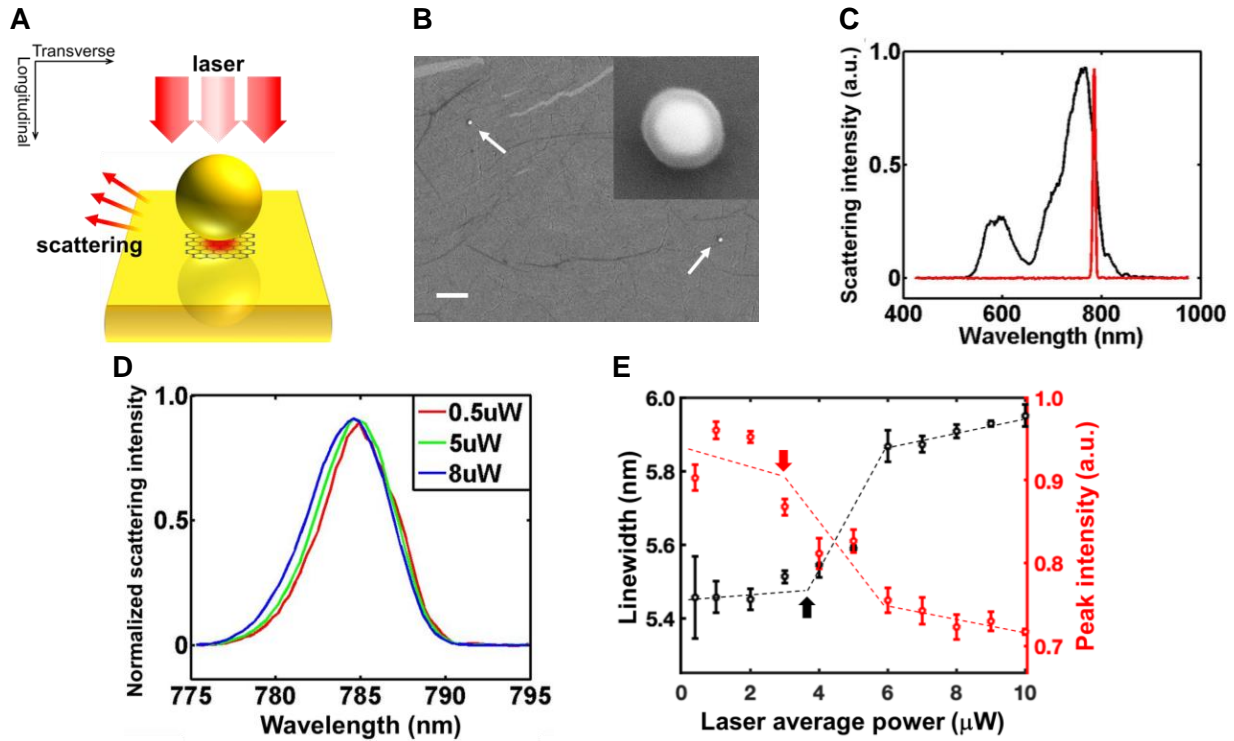

**Fig. 1. Experiment results for antennas embedded with a monolayer graphene sheet.** (**A**) Schematic of the experiment (not to scale). (**B**) An SEM image with arrows pointing to antennas, and an inset showing a zoom-in view of one antenna, scale bar 1 μm. (**C**) The scattering spectrum of an antenna under supercontinuum source illumination (in black), and the fs laser line (in red). (**D**) Normalized fs laser scattering spectra, after being smoothed with a window of 0.8 nm width. (**E**) FWHM linewidths and normalized peak intensities of fs laser scattering spectra with different laser powers. The dashed lines are eye guides. (C-E) were taken from the same antenna. The error bars in (E) were determined by measurement of laser reflection spectra off a glass slide, with three sequential measurements for each laser power.

In this experiment, the 2D material sheet is chosen for its high nonlinear interaction strength with electromagnetic fields. In particular, the large 3$^{rd}$-order optical nonlinearity of graphene is well recognized [42,43]. It becomes even larger when the pump, signal and idler wavelengths are close to each other due to a one-photon absorption saturation effect [21,44], which will further lower the switching threshold in our experiments since all of our wave-mixing components are contained in the same fs pulse. Recently, the $3^{rd}$-order interaction between nano-focused optical field and graphene has been experimentally proven to be much stronger than that of nonlocal plasma in gold [8,14,21], so we will ignore the contribution from gold to nonlinearity. In addition, recent reports show efficient high harmonic generation in 2D materials [45-48], which is essential for the formation of bistability as will be shown by our theory.

Figure 1, B-E, show the experiment results for when the 2D material is graphene. Each nanosphere was additionally covered with a 25 nm thick $Al_2O_3$ layer by atomic layer deposition, which tuned the antenna's localized surface plasmon resonance (LSPR) to align with the fs laser, and protected the nanostructure under laser impingement. A scanning electron microscopy (SEM) image of the antennas is shown by Fig. 1B. The linear LSPR scattering spectrum of one of the antennas is shown by Fig. 1C, which was obtained by focusing a supercontinuum source at normal incidence upon the antenna and collecting the scattered light from a side direction. A complete description of the optical measurement setup is in [49]. The scattering far-field of the longer-wavelength LSPR mode in Fig. 1C contained both longitudinal and transverse polarization components considerably (definition of directions in Fig. 1A), manifesting itself as a binding quadrupole (BQP) mode. The BQP mode contains a considerably enhanced transverse electric field in the gap so as to interact with the graphene sheet's large transverse nonlinearity [49].

Then we replaced the supercontinuum source with a fs fiber laser, which had a central wavelength of 785 nm, a repetition rate of 80 MHz and a linewidth of ~5 nm (which corresponds to a linewidth-limited pulse duration of ~200 fs). This laser line happened to align to the longer wavelength slope of the BQP spectrum, as in Fig. 1C. The laser scattering spectra were plotted in

Fig. 1D. It shows that, as the laser power increased, the scattering spectra broadened towards the LSPR peak direction, indicating an LSPR enhanced nonlinear effect. Meanwhile, there is a remarkable increase of intensity in the tails of the spectra on the LSPR peak side, indicating efficient generation of light at new frequencies.

The full-width at half-maximum (FWHM) linewidth and the peak intensity of the scattering spectrum *vs* the incident laser power are plotted in Fig. 1E. The spectral peak intensity has been normalized to the incident laser power. The most distinguishing feature of the experiment results is that both curves show a first turning from a relatively flat segment to an abrupt slope when the laser power increased over a certain value (labeled by thick arrows in Fig. 1E), and a second turning in the opposite manner at a higher laser power. In the theoretical model, we will attribute the first turning to transitioning from State 1 to State 2, and the second turning (flattening) to State 2 occupying most part of the pulse.

To make use of the even more intensive gap plasmons of the longitudinally polarized binding dipole (BDP) mode, we have replaced the monolayer graphene sheet with a monolayer $MoS_2$ sheet and performed the same testing. See [49] for a set of similar experiment results, in which the first turning-point laser power is as low as ~0.6 µW average or ~38 mW peak. In addition, see [49] for further analysis and control experiments to confirm that the abovementioned changes in fs laser scattering spectra were dominantly induced by optical nonlinearity rather than any linear effects or experimental uncertainty.

**Theory**—To understand the linewidth broadening and intensity dropping of fs laser scattering, and, in particular, the extraordinary turning phenomenon, we built a theoretical model to describe the nonlinear plasmon dynamics in the 2D-material-embedded antenna. We noticed that previous theories for bistability in the continuous-wave (CW) operation mode is not able to model the evolvement of the femtosecond pulse, for example, the almost uni-directional spectral broadening in Fig. 1B can't be reproduced by them. In addition, our model renders a universal quantitative threshold for switching in terms of FWM efficiency.

The working principle of the bistable switch is illustrated in Fig. 2. $E_L$ is used to indicate the antenna's near-field under a linear response to the incident laser, that is, when the nonlinear sheet conductivity of the 2D material $\sigma_{NL} = 0$. The total near-field, $E_{tot}$, generates a nonlinear sheet current density, $J_{NL}$, in the 2D material. $J_{NL}$ includes more than 3$^{rd}$-order nonlinearities. Then the radiation of $J_{NL}$ is enhanced by the Purcell effect of the antenna and adds an additional electric field, $E_{NL}$, to $E_{tot}$, forming a nonlinear feedback loop. The large nonlinearity of 2D materials and the large Purcell enhancement of NSoM together make this feedback strong, and the negative nonlinear current makes this feedback positive ($E_{NL}$ be in phase with $E_L$) when the laser and LSPR peak align with each other, so that the requirements for forming bistability are fulfilled.

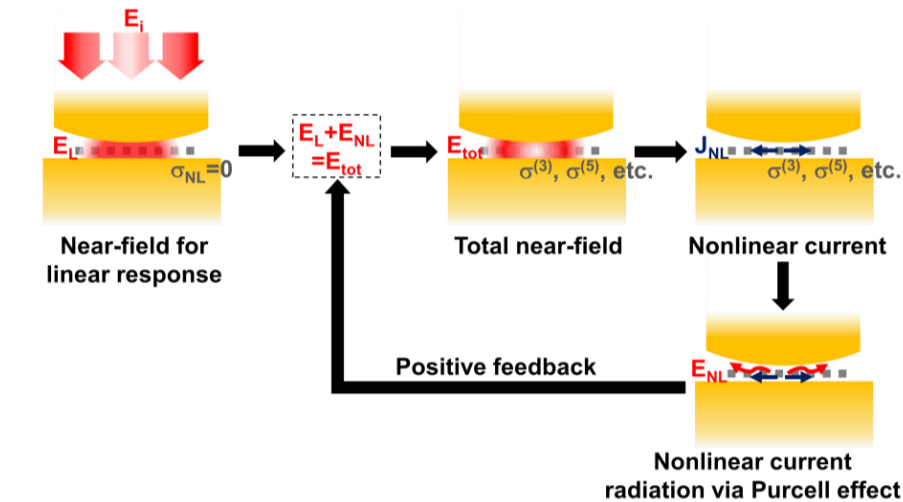

**Fig. 2. Working principle for a bistable plasmonic antenna.**

Following the flow diagram of Fig. 2, in the following, we calculate the nonlinear feedback and the dynamic switching. The model is based upon a few general assumptions, including the LSPR being Lorentzian and quasi-static, and $\sigma_{NL}$ being a negative real scalar. Therefore, it can be applied to various antenna geometries and gap interfaces.

First, we give the mathematical expression for $E_L$. Let the incident laser's electric field be $E_i(t)|E_i(1mW/\mu m^2)|\hat{i}e^{i\omega_1 t}$, where $\hat{i}$ is a unit vector for polarization, $\omega_1$ is the central frequency of incident laser, and $E_i(t)$ is the complex amplitude of a slowly varying envelop normalized by an intensity of $1mW/\mu m^2$. Let the linear LSPR response be a Lorentzian centered at $\omega_0$, with a plasmon oscillation lifetime of $\tau$ and a quality factor of Q. Let the LSPR mode profile be $\hat{E}(\vec{r})$, which is a complex vectorial amplitude of the near-field at position $\vec{r}$ when the antenna is linearly pumped by a CW laser at $\omega_0$ and $1mW/\mu m^2$. Then the antenna's linear near-field $E_L(t)\hat{E}(\vec{r})e^{i\omega_1 t}$ is determined by

$$E_L(\omega - \omega_1) = \frac{i\frac{\omega_0}{\tau}}{\omega_0^2 - \omega^2 + i\frac{\omega}{\tau}} E_i(\omega - \omega_1) \tag{1}$$

where $E(\omega)$ is Fourier transform of $E(t)$, bearing any subscripts.

Next, we calculate $E_{NL}$ from $J_{NL}$. Let the nonlinear sheet current density in the gap material be $\vec{J}_{NL}(t,\vec{r})e^{i\omega_1 t}$. Under a large Purcell enhancement, this current predominantly radiates into the LSPR mode, such that the nonlinear near-field takes the form of $E_{NL}(t)\hat{E}(\vec{r})e^{i\omega_1 t}$, which is determined by (see [49] for derivation)

$$E_{NL}(\omega - \omega_1) = -\frac{Q}{\omega_0 \int \varepsilon'(\vec{r})|\hat{E}(\vec{r})|^2 d^3\vec{r}} \frac{i\frac{\omega_0}{\tau}}{\omega_0^2 - \omega^2 + i\frac{\omega}{\tau}} \int_{gap} \vec{J}_{NL}(\omega - \omega_1, \vec{r}) \cdot \hat{E}^*(\vec{r}) d^2\vec{r} \tag{2}$$

where $\varepsilon'$ is the real part of dielectric constant. Here, the Lorentzian term stands for LSPR response, and the integral term is projection of the nonlinear current onto the LSPR mode profile which is further multiplied by the electric field enhancement factor (the Purcell effect).

Let the total near-field be $E_{tot}(t)\hat{E}(\vec{r})e^{i\omega_1 t}$. We have

$$E_{tot} = E_L + E_{NL}. \tag{3}$$

Next, we consider the situation where the nonlinearity is only 3$^{rd}$-order. By ignoring wave mixing at far away from $\omega_1$ (outside the LSPR spectral range), the 3$^{rd}$-order sheet current density is given by,

$$\vec{J}^{(3)}(t,\vec{r}) = \frac{3}{4}\sigma^{(3)}\left|E_{tot}(t)\hat{E}(\vec{r})\right|^2 E_{tot}(t)\hat{E}(\vec{r}) \tag{4}$$

where, for simplicity, we have assumed a scalar 3$^{rd}$-order sheet conductivity, $\sigma^{(3)}$. By substituting $\vec{J}^{(3)}$ for $\vec{J}_{NL}$ in Eq. (2), the near-field is

$$E_{NL}(\omega-\omega_1) = \frac{i\dfrac{\omega_0}{\tau}s}{\omega_0^2 - \omega^2 + i\dfrac{\omega}{\tau}} \mathcal{FT}[|E_{tot}(t)|^2 E_{tot}(t)]\bigg|_{\omega-\omega_1}, \text{ where } s = -\frac{3}{4}\frac{Q\sigma^{(3)}}{\omega_0}\frac{\int_{gap}\left|\hat{E}(\vec{r})\right|^4 d^2\vec{r}}{\int \varepsilon'(\vec{r})|\hat{E}(\vec{r})|^2 d^3\vec{r}}. \tag{5}$$

Finally, substituting Eq. (1,5) into Eq. (3), and assuming $\sigma^{(3)}$ as a negative real number, which is approximately true for graphene when the photon energy is above twice the chemical potential [44], we obtain the concluding equation for our model,

$$s^{1/2}E_{tot}(\omega-\omega_1) = \frac{i\dfrac{\omega_0}{\tau}}{\omega_0^2 - \omega^2 + i\dfrac{\omega}{\tau}}\{s^{1/2}E_i(\omega-\omega_1) + \mathcal{FT}[|s^{1/2}E_{tot}(t)|^2 s^{1/2}E_{tot}(t)]\bigg|_{\omega-\omega_1}\} \tag{6}$$

where field strength and material nonlinearity are combined into a dimensionless value, $s^{1/2}E$. Cascaded wave mixing is inherently included in this iteration-like equation for $s^{1/2}E_{tot}$.

It is important to point out that taking $\sigma^{(3)}$ as a negative real number makes $E_{NL}$ in phase with $E_L$ when $\omega_1=\omega_0$, such that maximum plasmon enhancement and optimum feedback phase are simultaneously achieved. In contrast, a number of previous proposals of plasmonic bistability were based on dielectric Kerr effect, in which the laser must be offset far from the resonance to compensate for a $\frac{\pi}{2}$ phase mismatch [38-40]. Further, we will show in the following this overlap between resonance and positive feedback demands higher-order nonlinearity be incorporated due to the requirement of energy conservation. It is interesting to note that such higher-order nonlinearity is implicitly included in the conventional equation for saturable absorption and the derived bistability thereof.

**Numerical Results**—First, to illustrate the necessity of incorporating higher-order nonlinearity, we plot $|E_i|$ vs $|E_{tot}|$ under a CW incident laser with $\omega_1=\omega_0$ in Fig. 3A, according to Eq. (6). With a constant $s$ value (pure 3$^{rd}$-order nonlinearity), the curve crosses the $|E_i|=0$ axis, indicating the existence of a near-field without any input. This violates the law of energy conservation. To correct this violation, $E_{NL}$ must saturate at high laser intensities. Therefore, we will adopt an arbitrary saturation relation in our model, $s = s_0(1-0.33 s_0 |E_{tot}^2|)$, which effectively incorporates 5$^{th}$-order nonlinearity into Eq. (6). Consequently, the $|E_i|$ vs $|E_{tot}|$ curve now contains two stable states (a stable state is such that a perturbation of the input leads to an increase or decrease of the response in the same direction). The arbitrary choice of $s$ saturation relation won't affect the following discussion on bistability and quantitative estimation of switching threshold.

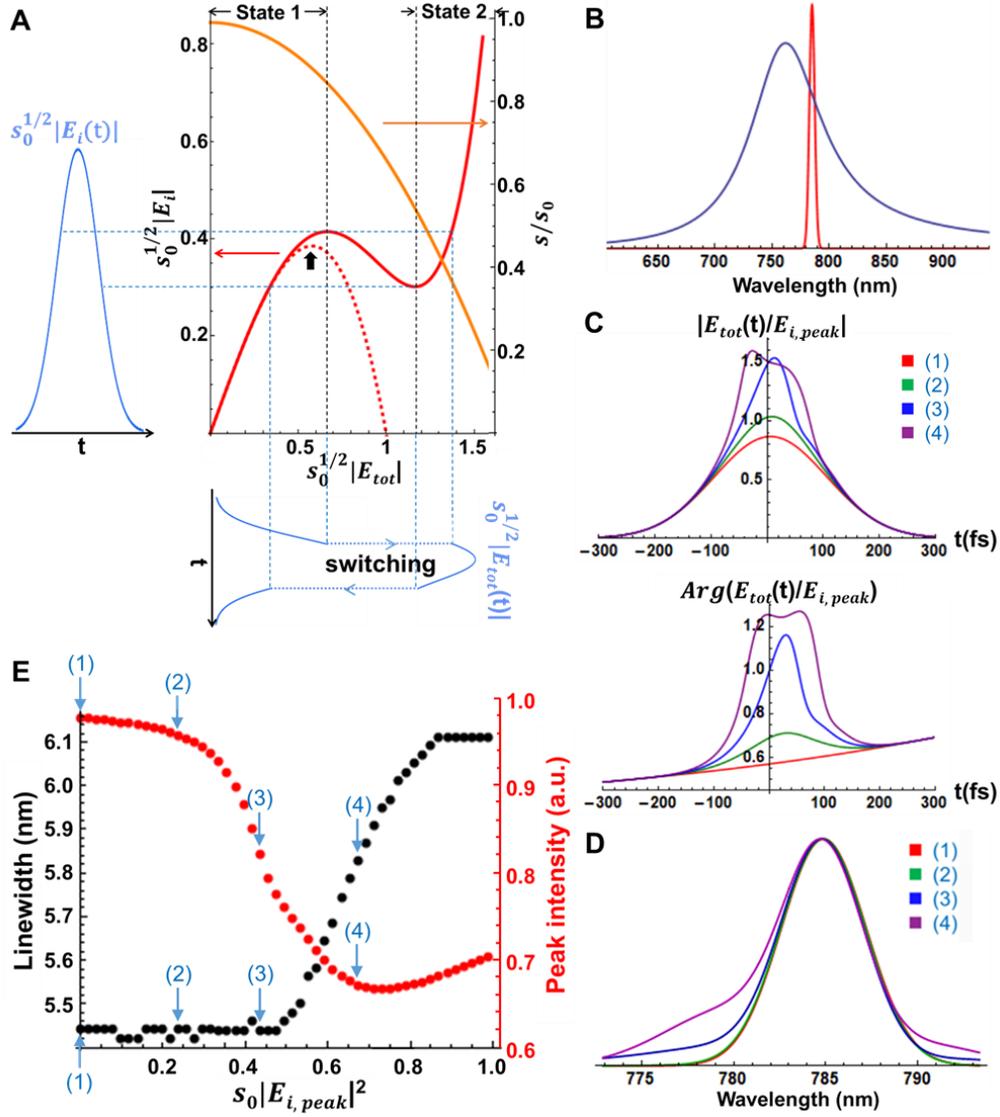

**Fig. 3. Modeling and results. For a CW incident laser at $\omega_0$: (A)** $|E_i|$ vs $|E_{tot}|$. The red dashed curve corresponds to a constant 3rd-order nonlinearity $s = s_0$, and the red solid curve corresponds to a saturated $s$ (shown by orange curve). The two blue curves schematically illustrate how a modulated laser switches the antenna. The thick black arrow points to an approximate switching threshold at $s_0|E_i|^2 = \dfrac{4}{27}$. **For a fs pulsed incident laser: (B)** Settings of LSPR intensity spectrum (in blue), and laser intensity spectrum (in red) for calculating laser scattering. **(C)** The amplitude

and phase of $E_{tot}(t)$, for $s_0|E_{i,peak}|^2 = 0, 0.24, 0.43, 0.67$. (**D**) The normalized far-field scattering spectra, for the same four cases. (**E**) The FWHM linewidths and normalized peak intensities of laser scattering spectra, with a charge retardation phase $\phi = -\frac{1}{3}\pi$.

If the CW laser is modulated so that $|E_i|$ increases from zero to above the plateau of State 1, the antenna will switch from State 1 to State 2 resulting in an abrupt change in $|E_{tot}|$, as illustrated by the blue curves. Another switching event happens during the decrease of $|E_i|$. The threshold condition for the first switching to happen can be approximately calculated from the plateau of the $s = s_0$ curve, which gives $(s_0|E_i|^2)_{th} = \frac{4}{27}$, as labeled by a thick black arrow. In [49], we show if the laser is sinusoidally modulated with a 100% modulation depth, a FWM efficiency of $\frac{1}{16}s_0^2|E_{i,peak}|^4$ will be obtained under 1$^{st}$-order perturbation assumption, where $|E_{i,peak}|$ is the peak value of $|E_i(t)|$. Therefore we have a universal switching threshold of *effective FWM efficiency ≈ 0.14%*. For laser frequencies away from $\omega_0$, the threshold will be higher. A theoretical estimate of the effective FWM efficiency for the graphene-embedded antenna is included in [49] to show that it is indeed reasonable to have reached above the switching threshold in our experiments.

In the following, we present the solution to Eq. (6) for pulsed pumping. The solving process was done by converting Eq. (6) to a time-domain 2$^{nd}$-order differential equation and letting $E_{tot}$ evolve with $t$ numerically. The plasmon lifetime, $\tau$, is set to 4 fs, and the incident laser pulse is set to a Gaussian with a full pulse width at 1/e-intensity of 200 fs. In addition, the laser line aligns with the side slope of LSPR spectrum, as shown by Fig. 3B. These settings have been chosen to match those of the experiments. Fig. 3C shows the calculated $|E_{tot}(t)|$, whose shape is nearly Gaussian at low laser intensities, but contains fast rises and falls (state transitions) at high laser intensities. In addition, the phase of $E_{tot}$ is also shown to be significantly modified by state transitioning.

To calculate the far-field scattering spectra, it should be noticed that the charge retardation effect of the antenna introduces an extra phase difference, $\phi$, between the far-field radiation of the linearly scattered light and the far-field radiation of the nonlinear gap plasmons (mediated by the antenna) in addition to their near-field phase difference. Therefore, the linearly and nonlinearly scattered far-field components interfere by a total phase difference of $Arg(E_L) - Arg(E_{NL}) + \phi$. Further explanation of the charge retardation phase and interference is in [49]. Using $\phi = -\frac{1}{3}\pi$, the scattering spectra of the pulsed laser are plotted in Fig. 3D.

The linewidth and peak intensity of the scattering spectrum *vs* the incident laser intensity are plotted in Fig. 3E. The data points corresponding to the four cases in Fig. 3C are labeled, illustrating that the time-domain fast transitioning between State 1 and 2 leads to the broadening of scattering linewidth, and the (destructive) interference between linear and nonlinear far-field components leads to the drop of scattering intensity. As State 2 occupies most part of the pulse at the highest laser intensities, the broadening and dropping trends flatten and bend over.

Comparing Fig. 1 and 3, our theory has nicely reproduced all of the important features of the fs laser scattering experiment results, including an over 10% linewidth broadening towards the LSPR peak direction, a remarkable rise of the shorter wavelength spectral tail, a slight blueshift of the spectral peak, and abrupt turnings and flattenings in the linewidth and intensity *vs* laser power curves. To achieve this nice match between theory and experiment, the only fitting parameter in our calculation is the charge retardation phase. As an aside, the experimental laser pulses were not perfectly Gaussian, as indicated by the asymmetric line-shape in Fig. 1D, which could explain the difference between the shapes of the spectral tails in Fig. 1D and 3D.

**Discussion**—We have found extraordinary changes of the scattering spectra as we pumped 2D-material-embedded NSoM resonators with a fs laser. By building a plasmon dynamics model, we have identified the spectral changes to be the sign of transitioning between different plasmon states. The model nicely reproduced the experiment results, and enabled us to infer bistability without conducting an ultrafast measurement of the hysteresis loop [50]. In addition, the model tells the difference between the switching mechanism of our devices and that of dielectric Kerr effect, including the role of higher-order nonlinearity.

Being energy efficient, nanometer footprinted and ultrafast, we envisage 2D-material-assisted gap plasmon bistable switch to be a promising candidate for memory and calculation units in future integrated circuits. The operating laser power for the MoS$_2$-embedded NSoM was around a thousandth of those for previous nanoplasmonic nonlinear devices, and its flip-flop energy was less than 10 fJ/bit. Further, corresponding to the universal threshold of effective FWM efficiency≈0.14% is a threshold laser power approximately proportional to $\sigma^{(3)-1}Q^{-2}$ (or $s_0^{-1}$). Since $\sigma^{(3)}$ is nonlocal and will increase much by nano-structuring the 2D material [21,46], and Q will improve by using plasmonic materials with low ohmic losses [51], in principle, the operating laser power could be reduced by several orders of magnitude from the experiments in this work so as to work under a CW bias. It thus shows an unprecedented potential to achieve a memory bandwidth per unit area that is orders of magnitude larger than electronic SRAMs and dielectric optical bistable resonators [52-54].

**ACKNOWLEDGMENTS:** We thank Shouwu Guo's group at Shanghai Jiao Tong University (SJTU), Chengwei Qiu from National University of Singapore, Zhaowei Liu from University of California at San Diego, Yuanbo Zhang's group at Fudan University, and Wenqi Zhu from University of Maryland for helpful discussion or technical assistance. Sample fabrication was supported by the Center for Advanced Electronic Materials and Devices of SJTU. Numerical simulation was supported by the Center for High Performance Computing of SJTU. This work is funded by the National Natural Science Foundation of China (grants 11574207 and 61975253), the Natural Science Foundation of Shanghai (grant 18ZR1421600), and the National Infrastructures for Translational Medicine (Shanghai).

**AUTHOR CONTRIBUTIONS:** T.Y. proposed and supervised the project, interpreted the experiment results, and developed the theoretical model. T.Y., H.Y. and J.L. built the experiment setup and developed the experiment methods. T.Y., X.W., C.C. and X.K. completed the experiment design, finished the experiments, and analyzed the data. X.W. collected the experiment data. Y.M. performed FDTD simulations. All authors participated in discussion and preparation of the manuscript. T.Y., H.Y. and X.W. are co-first authors.

* tianyang@sjtu.edu.cn

# Supplemental Material for

# 2D-Material-Assisted Bistable Switching of Gap Plasmons Disclosed By Femtosecond Pulse Scattering Spectra

Tian Yang*, Hui Yi, Xiaodan Wang, Yichen Miao, Cheng Chen, Jing Long, Xiangyang Kong, Lin Wu.

Correspondence to: tianyang@sjtu.edu.cn

**This PDF file includes:**

1. Experiment methods
2. Experiment results for $MoS_2$-embedded antennas
3. Experiment results for LSPR linear scattering spectra and polarization
4. Additional analysis and control experiments to confirm the validity of experiment results
5. Proof of equation (2)
6. Derivation of FWM efficiency under $1^{st}$-order perturbation assumption
7. Theoretical estimate of effective FWM efficiency for the graphene-embedded antenna
8. Simulation and discussion for charge retardation and far-field interference

## 1. Experiment methods

Sample Preparation

In the graphene-embedded antennas, the mirror plane was a 200 nm thick gold film on a mica substrate, which had been treated with a hydrogen flame to achieve an atomically flat gold surface. Then the sample was rinsed with acetone, isopropanol and ultra-purified water, and dried under a stream of nitrogen. Next a monolayer graphene sheet sandwiched between a water-soluble polymer layer and a polymethyl methacrylate (PMMA) layer was wet transferred to the gold plane. Next the sample was dried in ambient conditions for 24 hours, immersed in acetone for 0.5 hours to remove PMMA, and immersed in 1:1 isopropanol to remove acetone. Last, a droplet of 100 nm diameter gold nanosphere colloid with a concentration of $10^9$/mL was drop casted onto the sample. After 45 minutes, the sample was dried under a stream of nitrogen to remove the colloid solution and finish the sample preparation.

To prepare the $MoS_2$-embedded antennas, the atomically flat gold plane was treated by deep ultraviolet light to improve surface affinity, and the $MoS_2$ sheet was transferred from a sapphire substrate (done by the company Six Carbon). The rest of the procedures were the same as above.

Optical measurement setup

A schematic of the home-built optical system for LSPR, laser scattering and SERS measurements is shown in Fig. S1. The light source was one of the following: a super continuum source, a fs fiber laser and a He-Ne CW laser at 632.8 nm. The light source was first collimated and expanded to a 4 mm diameter beam, then focused onto the sample through a 100× objective with a numerical aperture (NA) of 0.9, resulting in a focal spot of about 0.5 μm diameter on the sample. The scattered light of the supercontinuum source and the fiber laser off the sample was collected from a side direction using a lens with a NA of 0.15. This collecting lens focused the scattered light into a fiber-bundle, which was connected to a monochromator and a charge coupled device (CCD). The CCD was in fact an electron-multiplying (EM) CCD, but high EM gain would bring considerable errors to the measurements and had better be not used. The combination of sample scattering, collecting lens and fiber-bundle forms a confocal scheme that effectively suppressed collection of stray light. In addition, the fs fiber laser was filtered by a 10 nm bandwidth band-pass filter before being focused onto the sample, in order to clean its spectral profile. On the other hand, using the He-Ne laser, Raman scattering off the sample was collected by the same 100× objective, and a long-pass filter was used to block the laser light from entering the monochromator. The power of the light sources was tuned by neutral density filters.

In addition, a camera lens, a CMOS camera and a lamp were used in combination with the 100× objective to form a microscope, which were used to find the antennas and align them to the centers of the optical focal spots.

## 2. Experiment results for MoS$_2$-embedded antennas

By inserting a monolayer MoS$_2$ sheet in the nanosphere-on-mirror antennas' gaps and performing the same testing as for the graphene-embedded ones, a similar set of diagrams for the experiment results were obtained as shown by Fig. S2, showing exactly the same critical features of state transitioning, except for the flattening and bending-over parts which could be beyond the safe laser power range. Here the gold nanospheres were not covered with an Al$_2$O$_3$ layer (Fig. S2A), since their LSPR naturally matched the fs laser (Fig. S2B). The scattering far-field of the longer wavelength LSPR mode was longitudinally polarized, manifesting itself to be a BDP mode (See Section 3 of Supplemental Material). Due to the lack of Al$_2$O$_3$ protection layer, the average incident laser power was limited to 1 µW, which corresponds to a peak instantaneous power of 62.5 mW assuming a linewidth limited pulse duration of 200 fs. Above this power, the LSPR spectra could be irreversibly modified by laser illumination according to our experience. A turning-point laser power as low as ~0.6 µW average or ~38 mW peak can be extracted from Fig. S2D. The fundamental mechanism that renders high nonlinearity in the metal-MoS$_2$-metal structure is a subject for future investigation.

## 3. LSPR linear scattering: spectra and polarizations

The LSPR scattering spectra of multiple graphene-embedded antennas, before and after Al$_2$O$_3$ coating, are shown in Fig. S3, A and B. Before Al$_2$O$_3$ coating, a transverse mode near about 580 nm and a BQP mode near about 700 nm were observed. The BQP mode of some antennas split into two spectral peaks, which was attributed to coupling between antennas and gap plasmons [55]. Otherwise, the variation of the LSPR spectra was small between different antennas. After Al$_2$O$_3$ coating, some antennas showed redshifts of the BQP mode, and the others showed blueshifts where the BQP mode almost merged into the transverse mode. While the redshifts follow what normally happens for LSPR refractive index sensors, the blueshifts indicate that $Al_2O_3$ was not only coated around the nanospheres but also filled into some of the junction gaps and enlarged the gap distances. We picked those antennas whose redshifted BQP spectra aligned with the fs laser to measure the nonlinear scattering behaviors.

The LSPR scattering spectra of multiple $MoS_2$-embedded antennas are plotted in Fig. S3C, showing a BDP mode near about 780 nm, and a small variation between each other.

In addition, we measured the longitudinal and transverse polarization components of the LSPR scattering spectra for the antennas (Fig. S4), by inserting a polarizer in the confocal light path for scattering collection (blue path in Fig. S1). The figure shows that, near 785 nm, the scattered far-field contained both longitudinal and transverse components considerably for the graphene-embedded antenna, while it was predominantly longitudinal for the $MoS_2$-embedded antenna. Therefore, they are assigned as BQP and BDP modes, whose gap plasmons also contain both longitudinal and transverse components, or are predominantly longitudinal, respectively [56].

## 4. Additional analysis and control experiments

The following analysis and control experiments further confirm that the experimental linewidth and intensity trends as reported in this paper were dominantly induced by optical nonlinearity rather than any linear effects or experimental uncertainty.

First, there is a remarkable increase of the shorter wavelength tails of the normalized laser line spectra, for example, as marked by a dashed line in Fig. S2C, while the longer wavelength halves of the laser line remain almost unchanged. None of the currently known linear effects of plasmonic antennas is likely to distort the linear LSPR response so steeply within the laser line wavelength range for this to happen, including refractive index change [57], antenna morphology change [58], picocavity formation [59] and memristive atom movement [60]. A schematic of the required (and impossible) LSPR distortion is shown in Fig. S5.

Second, we checked LSPR and surface-enhanced Raman scattering (SERS) of the samples before and after the testing, to confirm that they were not irreversibly deformed or degraded by laser impingement. The LSPR scattering spectra under supercontinuum source illumination are shown by Fig. S6, A and B. The SERS spectra under He-Ne laser pumping are shown by Fig. S6C. No obvious changes have been found before and after the testing. We limited the fs laser average power to below 1 μW for the MoS$_2$-emdded antennas, and to below 10 μW for the graphene-embedded ones with an Al$_2$O$_3$ protection layer.

Third, we tested fifteen antennas in total, among which six showed 0.5 nm or more linewidth broadening. All of the six antennas showed the abrupt "turn-on" characteristic in their linewidth *vs* laser power curves, showing a good reproducibility of the experiment results.

Fourth, we have conducted a theoretical estimate on the effective FWM efficiency for the graphene-embedded antenna, and showed that it is indeed reasonable to have reached the switching threshold under our experiment conditions (See Section 7 of Supplemental Material).

## 5. Proof of equation (2)

First, consider the situation in which the nonlinear electromagnetic fields are monochromatic and have a frequency of $\omega_0$. Then, the radiation plus absorption power from the nonlinear near-field of the antenna, $E_{NL}\hat{E}(\vec{r})e^{i\omega_0 t}$, is (note that at $\omega_0$, $|E_L|=1$ under 1 mW/μm$^2$ pumping)

$$\text{nonlinear extinction power} = 1\frac{mW}{\mu m^2}\sigma_{ext}|E_{NL}|^2 \tag{S1}$$

where $\sigma_{ext}$ is the extinction cross section of the antenna at $\omega_0$. At the same time, energy flows from the nonlinear current source, $\vec{J}_{NL}$, into the nonlinear near-field by a power of

$$\text{nonlinear current output power} = -\frac{1}{2}\int_{gap}\vec{J}_{NL}(\vec{r})\cdot E_{NL}^*\hat{E}^*(\vec{r})d^2\vec{r}. \tag{S2}$$

Note that the projection of $\vec{J}_{NL}(\vec{r})$ on to the LSPR mode profile, which is the effective source current, is $\int_{gap}\vec{J}_{NL}(\vec{r})\cdot\frac{\hat{E}^*(\vec{r})}{\sqrt{\int_{gap}|\hat{E}(\vec{r})|^2 d^2\vec{r}}}d^2\vec{r}\frac{\hat{E}(\vec{r})}{\sqrt{\int_{gap}|\hat{E}(\vec{r})|^2 d^2\vec{r}}}$. It must leads its radiation field, $E_{NL}\hat{E}(\vec{r})$, by a phase of $\pi$. This is based on the general rule that the power flow direction between a source and its radiation be from the former to the latter at any time moments when the system is on its central resonance frequency. This explains why the right side of equation (S2) gives a positive real number. Then by equaling equation (S1) and (S2), that is, the sum of ohmic and radiation losses of the nonlinear near-field equals the output power of the nonlinear current source, we obtain

$$E_{NL} = -\frac{1}{2} \frac{1}{1\frac{mW}{\mu m^2} \sigma_{ext}} \int_{gap} \vec{J}_{NL} \cdot \hat{E}^*(\vec{r}) d^2\vec{r}. \tag{S3}$$

By substituting the following equation (S4) into (S3) and adding the off-resonance Lorentzian term, equation (2) is proved.

$$1\frac{mW}{\mu m^2} \sigma_{ext} = \frac{1}{\tau}[\text{total energy in } \hat{E}(\vec{r})] = \frac{1}{Q/\omega_0} \frac{1}{2} \int \varepsilon'(\vec{r}) |\hat{E}(\vec{r})|^2 d^3\vec{r}. \tag{S4}$$

## 6. Derivation of FWM efficiency under 1st-order perturbation assumption

In the following, we relate the value of $s$ to an *effective FWM efficiency*. Assume the incident field is a CW laser modulated by a sinusoidal function, in which the laser frequency is $\omega_0$, the field modulation frequency is $\Delta\omega$ ($\ll 1/\tau$), and the modulation depth is 100%. The intensity of the laser, $I$ (mW/μm$^2$), is related to the incident electric field amplitude $E_{i0}$ by (according to the definition of $E_i(t)$ in the main text)

$$E_{i0} = I^{1/2}. \tag{S5}$$

The incident field is thus modulated to be

$$E_i(t)e^{i\omega_0 t} = \frac{1}{2} E_{i0}(e^{i\Delta\omega t} + e^{-i\Delta\omega t})e^{i\omega_0 t}. \tag{S6}$$

Assume FWM is a 1st-order perturbation, which means it is so weak that we will rewrite equation (3) as $E_{tot} = E_L$. In addition, by equation (1), we have $E_L = E_i$. Therefore, we have $E_{tot} = E_i$. Then according to equation (5), the 3rd-order nonlinear near-field is

$$\begin{aligned}E_{NL}^{(3)}(t)e^{i\omega_0 t} &= 2^{-3} E_{i0}^3 s_0 \left| e^{i\Delta\omega t} + e^{-i\Delta\omega t} \right|^2 (e^{i\Delta\omega t} + e^{-i\Delta\omega t})e^{i\omega_0 t} \\ &= 2^{-3} E_{i0}^3 s_0 (3e^{i\Delta\omega t} + 3e^{-i\Delta\omega t} + e^{i3\Delta\omega t} + e^{-i3\Delta\omega t})e^{i\omega_0 t}\end{aligned} \tag{S7}$$

in which the FWM idlers are $2^{-3} E_{i0}^3 s_0 (e^{i3\Delta\omega t} + e^{-i3\Delta\omega t})e^{i\omega_0 t}$, which contain two new frequency components with the same $E^2$-intensity of $2^{-6} s_0^2 E_{i0}^6$. Since the linear near-field $E_L(t)e^{i\omega_0 t}$ has an average $E^2$-intensity of $2^{-1} E_{i0}^2$, the power ratio between idler scattering and linear scattering is

$$\textit{effective FWM efficiency} = \frac{1}{16} s_0^2 E_{i0}^4, \text{ or } \frac{1}{16} s_0^2 I^2. \tag{S8}$$

In the main text, we use $E_{i,peak}$ in place of $E_{i0}$.

The above imaginary experiment, in which a sinusoidal slowly modulated CW laser is used to pump the antenna, gives an *effective FWM efficiency* for each laser intensity $I$. It tells the relation between bistable switching and FWM, which is, when a threshold *effective FWM efficiency* is reached, bistable switching will happen (for laser frequency of $\omega_0$).

## 7. Theoretical estimate of effective FWM efficiency for the graphene-embedded antenna

In the following, we give a theoretical estimate of the effective FWM efficiency (as defined in Section 6 of Supplemental Material) for the graphene-embedded antenna experiment, and show that it is reasonable to have gone beyond the switching threshold under our experiment conditions. Due to the assumption of weak FWM in the definition of effective FWM efficiency, the following calculation will be based on the linear gap plasmon enhancement factor, $EH$. The same imaginary FWM experiment as defined by equation (S5-S8) is considered.

Let the electric field amplitude of the focal spot of the CW laser after the focusing objective be $E_0$ ($= E_{i0} |E_i(1mW/\mu m^2)|$). Then the electric field amplitude at the center (or peak) of the CW gap plasmon hotspot is $EH \times E_0$. Point dipole radiation from the gap center is enhanced by the same factor of $EH$ due to a reversible relation [61].

The power intensity of the focused CW laser, $I$, is related to $E_0$ by

$$I = \frac{1}{2}\varepsilon_0 c E_0^2 \qquad (S9)$$

where $c$ is the speed of light in vacuum, and $\varepsilon_0$ is the permittivity of vacuum. Though equation (S9) is only valid for planewaves, we use it for the purpose of a rough estimate here.

After modulation, the dipole moment of each of the two FWM idlers in the 2D material is

$$p^{(3)} = \frac{3}{4}\varepsilon_0 \chi^{(3)} EH^3 (\frac{1}{2}E_0)^3 V . \qquad (S10)$$

where $V$ is the gap plasmon hotspot's volume. Note that the volume of $E^8$ should be applied here since FWM experiences quadruple enhancement. Taking a $E^2$-hotspot diameter of $\sqrt{dg}$, where $d$ is the nanosphere diameter and $g$ is the height of the gap [62], and assuming the hotspot intensity distribution profile as Gaussian for simplicity, the $E^8$-$V$ is estimated to be $V = \frac{\pi}{16} dg^2$.

Through the Purcell effect, the radiation of $p^{(3)}$ into the upper half space is mediated by the antenna and enhanced by a factor of $EH^2$ compared to free-space radiation (here we have assumed that, in the upper half space, the antenna LSPR mode's far-field radiation profile is approximately the same as the far-field radiation profile of $p^{(3)}$ in free space). So the radiation power (or nonlinear scattering power) of each idler is given by (assuming $p^{(3)}$ as a point dipole)

$$P_{sca}^{(3)} = \frac{(\omega_0)^4}{24\pi\varepsilon_0 c^3}[p^{(3)}]^2 EH^2. \tag{S11}$$

Then we have

$$\text{effective FWM efficiency} = \frac{2P_{sca}^{(3)}}{\frac{1}{2}\sigma_{sca}I} = \frac{3\pi\omega_0^4}{2^{16}c^6\varepsilon_0^2} \frac{[\chi^{(3)}]^2 d^2 g^4}{\sigma_{sca}} EH^8 I^2 \tag{S12}$$

where the linear scattering power of the modulated laser is given by $\frac{1}{2}\sigma_{sca}I$, $\sigma_{sca}$ being the LSPR linear scattering cross section.

Let's assume the following approximate parameters for an antenna embedded with a monolayer graphene sheet: LSPR resonance at 785 nm, $\chi^{(3)} = 10^{-16} \, m^2/V^2$ [42,43], $d=100$ nm, $g = 0.5$ nm, $\sigma_{sca} = 0.2 \, \mu m^2$, and $EH = 100$. Then, according to equation (S12), focusing a 45 mW peak laser power to a 0.5 μm focal spot upon the antenna corresponds to an effective FWM efficiency of ~0.14%. Therefore, in theory, it is indeed reasonable to have reached above the switching threshold in our experiments.

## 8. Charge retardation (CR) and far-field interference

Figure S7, A-D, show the simulation results for a point dipole source placed in the gap of an NPoM antenna. The antenna consists of a 100 nm gold nanosphere on a flat gold substrate, with a 2 nm high gap filled with a gap layer material (Fig. S7A). The gap layer has a refractive index of 2.9, following the longitudinal refractive index of graphene [63]. The point dipole oscillates along the longitudinal direction, with its Purcell factor spectrum show in Fig. S7B. When the dipole oscillates at 652 nm (peak of a BDP mode), the amplitude and phase profiles of the electric field are shown in Fig. S7, C and D. Fig. S7C tells that the surface charges mainly accumulate near the top and bottom of the nanosphere, since surface charge density can be approximately estimated as $\varepsilon|E|$ on the nanosphere's surface boundary by Gauss' Law. At the same time, Fig. S7D shows that the phase of $E_z$ continuously decreases from the bottom to the top, showing a strong CR effect.

In Fig. S7E, the surface charge distribution of the nanosphere and its phase profile are schematically illustrated by simplified plots for two different situations. The first situation is for linear scattering, where a laser is incident from the top. The second situation is for nonlinear gap plasmon radiation, where the source is $J_{NL}$ in the gap. The phases of surface charges at the top and bottom of the nanosphere and their mirror images are labeled, which can also be directly related to the phase of $\varepsilon E$ by Gauss' Law. Since the far-field radiation from the bottom-of-nanosphere charges and gap-enclosed charges are largely cancelled by their mirror images, which are very close to each other, the total far-field radiation is dominantly generated by the top-of-nanosphere charges and their mirror image. Therefore, we define the CR phase to be how much the top-of-nanosphere charges lag behind the bottom-of-nanosphere charges, the latter being directly related to the gap plasmons. For the linear scattering situation, since the nanosphere is immersed in the laser focal spot, we set the CR phase to $\phi_2 \approx 0$. For the gap plasmon radiation situation, according to Fig. S7D, we roughly set the CR phase $\phi_1 \approx \frac{5}{3}\pi$. Consequently, the CR induced extra phase lag between linear and nonlinear far-field radiations is $\phi = \phi_1 - \phi_2 \approx -\frac{\pi}{3}$. Then, by taking the total far-field to be proportional to $E_L + E_{NL}e^{-i\phi}$, we have obtained Fig. 3, D and E.

The above simulations were conducted with the finite-difference time-domain (FDTD) method by using a commercial software (Lumerical). The electrical permittivity of gold was obtained from [64]. The simulation mesh's grid size was 0.2 nm inside and near the gap, and 0.8 nm elsewhere. Although this is a much simplified and very rough estimation of the CR effect, and the modeled structure wasn't exactly the same as that in the experiment, it is good enough to show the strong CR effect. After all, due to the large variation of the 2D material permittivity values as reported in the literature, and the quantum nonlocality and quantum tunneling effects, it is not simple to perform an accurate FDTD simulation of the monolayer-graphene-embedded NSoM.

**Fig. S1.**

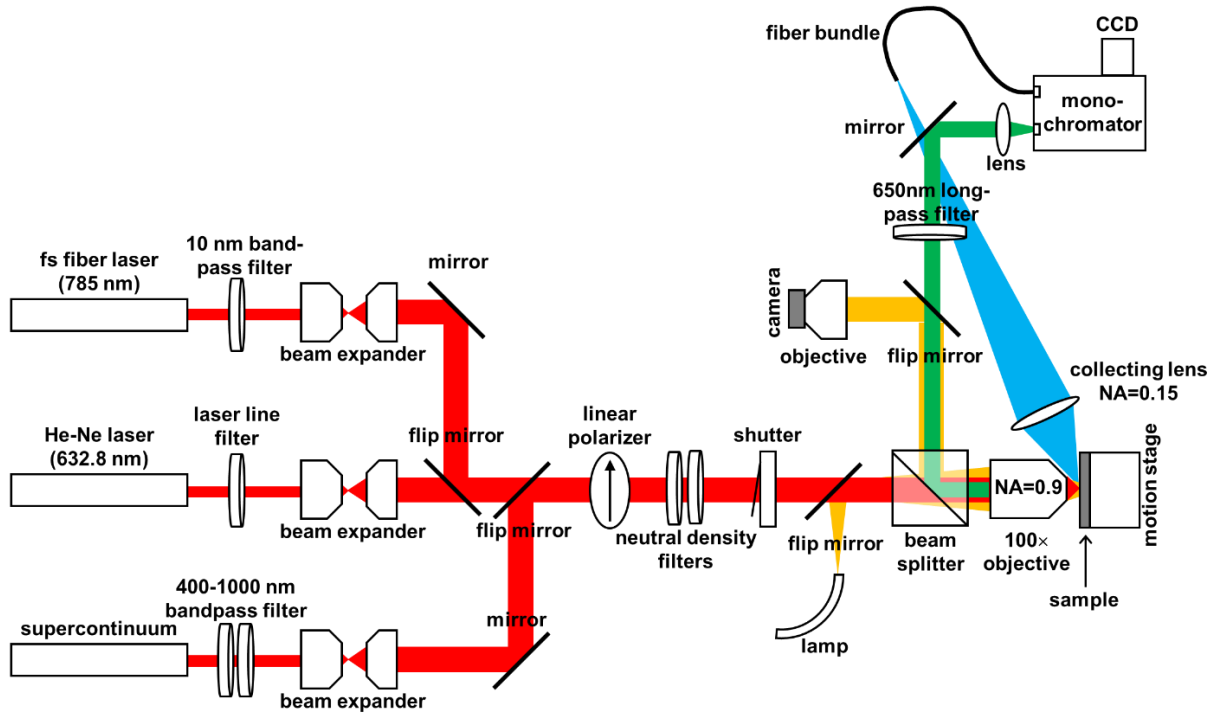

**Optical measurement setup.** Red: pump; yellow: optical microscopy; green: Raman spectroscopy; blue: scattering spectroscopy.

**Fig. S2.**

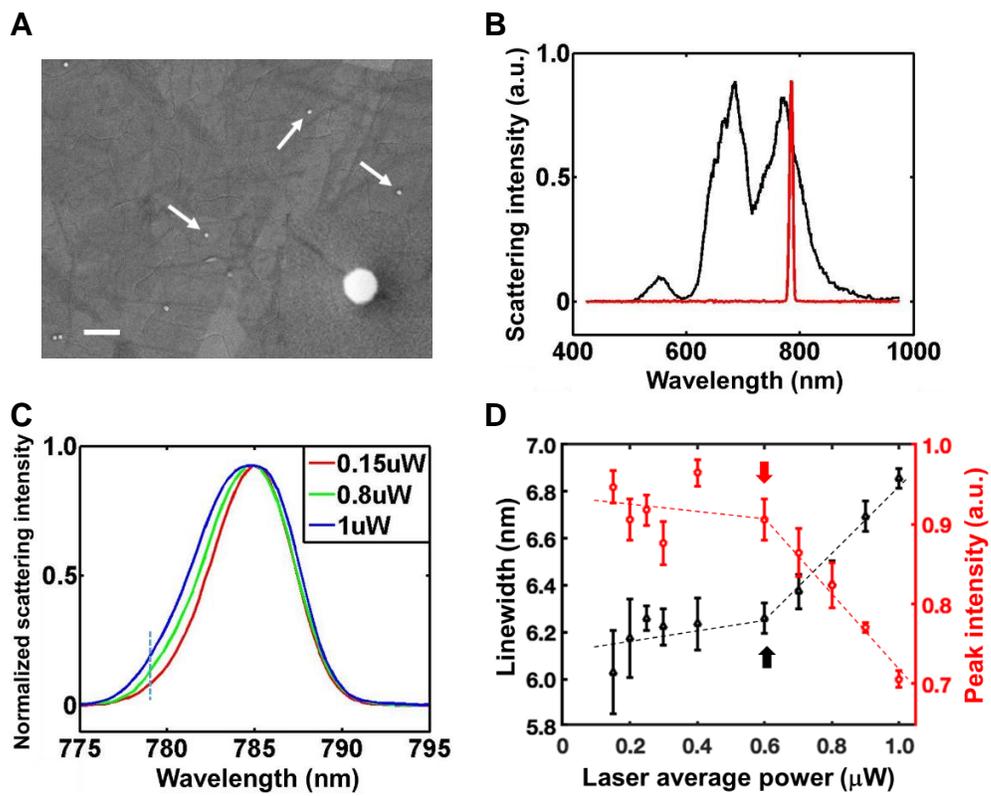

**Experiment results for antennas embedded with a monolayer MoS₂ sheet.** Same diagrams as Fig. 1, B-E, by replacing graphene with MoS₂.

**Fig. S3.**

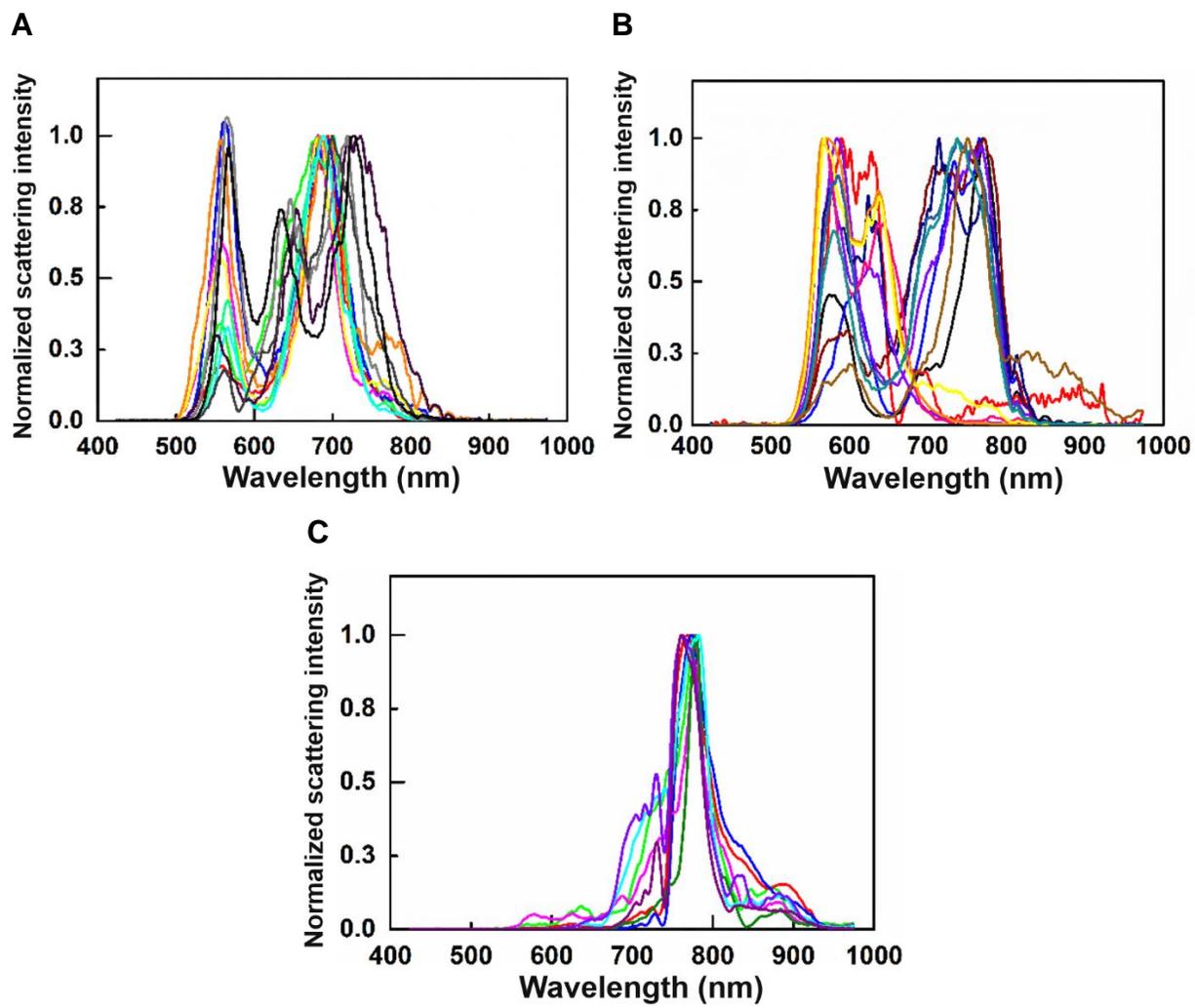

**LSPR spectra of multiple antennas.** (**A**) Twelve graphene-embedded antennas before $Al_2O_3$ coating. (**B**) Another twelve graphene-embedded antennas after $Al_2O_3$ coating. (**C**) Ten $MoS_2$-embedded antennas.

**Fig. S4.**

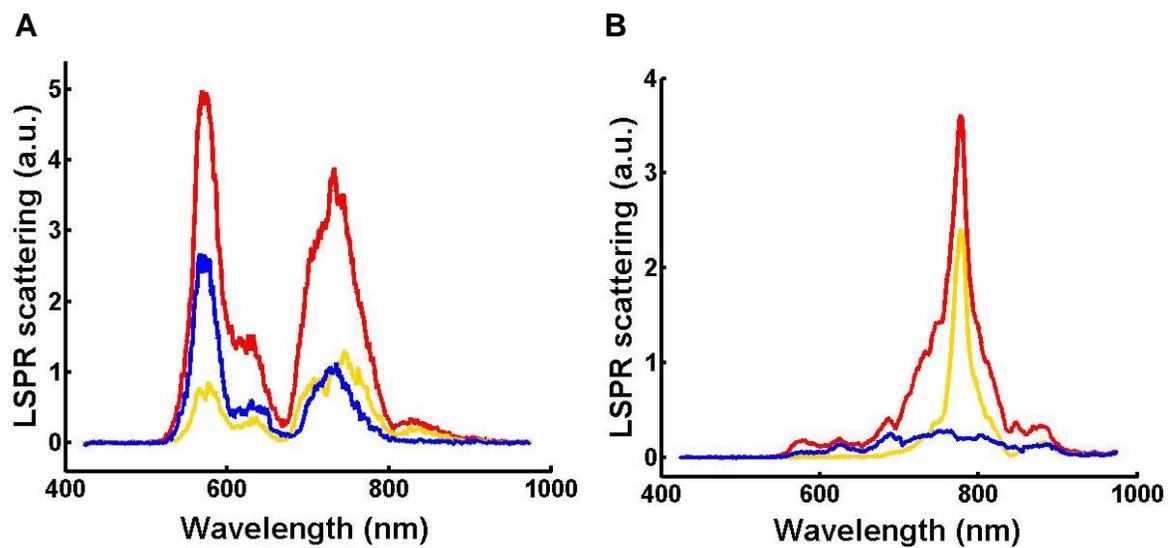

**Polarization of LSPR scattering.** (**A**) A graphene-embedded antenna. (**B**) A $MoS_2$-embedded antenna. Red: no polarizer; yellow: through a longitudinal polarizer; blue: through a transverse polarizer.

**Fig. S5.**

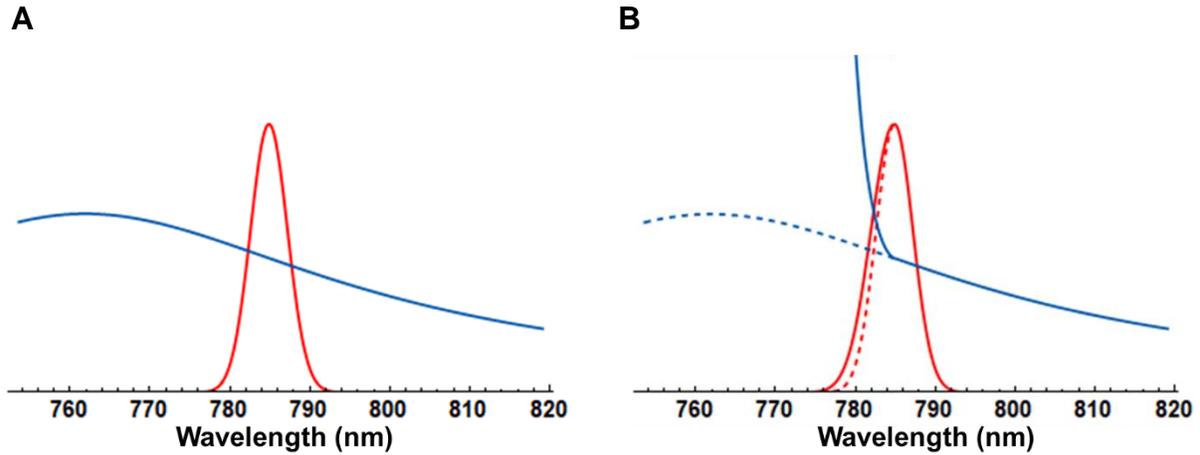

**Broadening and tail rising of fs laser scattering spectra can't be attributed to linear changes of LSPR.** To raise the spectral tail of the normalized scattered laser line by as much as in our experiments by purely linear effects, the LSPR linear scattering spectrum must be distorted steeply within the laser line wavelength range. Here is an illustration for what such a distortion should look like and how impossible it would be so. (**A**) A zoom-in view of the LSPR linear scattering spectrum (blue) and the laser scattering spectrum under a low laser intensity (red), using the same calculation settings as for Fig. 3. (**B**) To turn the low intensity laser scattering line (red dashed, same as in (A)) to the broadened one (red solid) by changing the LSPR response, the LSPR spectrum should change from the original (blue dashed, same as in (A)) to the steeply distorted (blue solid).

**Fig. S6.**

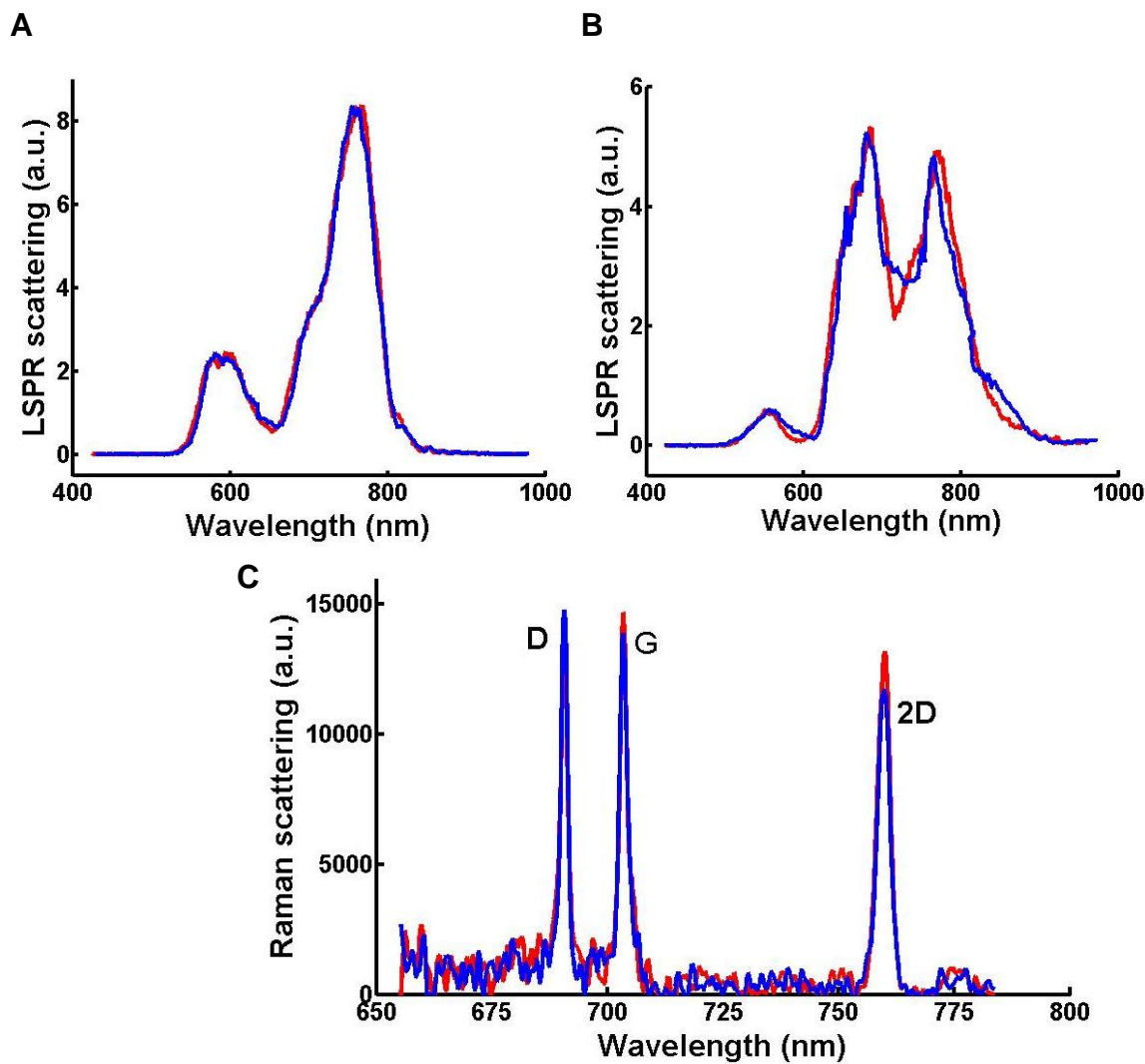

**LSPR and SERS spectra before and after fs laser pumping.** (**A,B**) LSPR scattering spectra under supercontinuum source illumination for the graphene-embedded antenna (A) and the $MoS_2$-embedded antenna (B) before (red) and after (blue) the fs laser scattering measurements. (**C**) SERS spectra for the graphene-embedded antenna before (red) and after (blue) the fs laser scattering measurements, with the feature Raman bands labeled. A 50 µW He-Ne laser and a 4 second integration time were used for the Raman spectroscopy measurements. The antennas whose measurement results are shown here are the same as the ones in Fig. 1, C-E and Fig. S2.

**Fig. S7.**

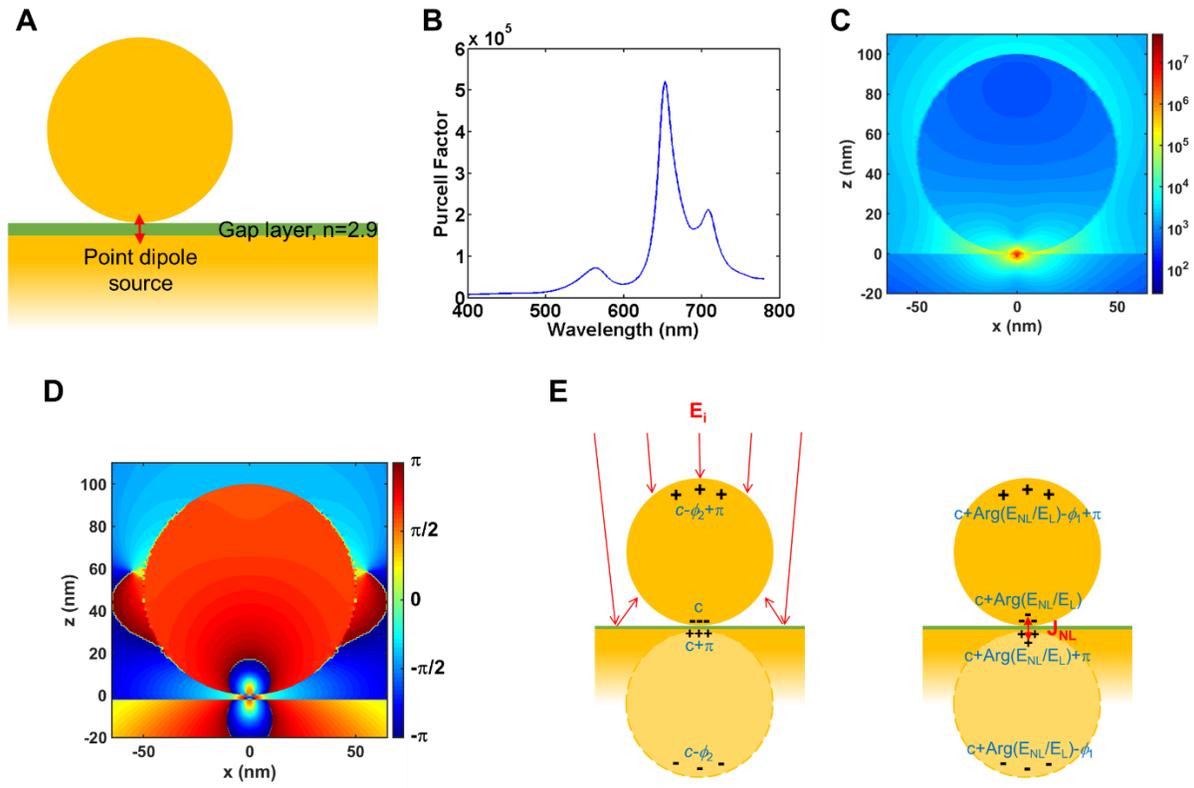

**Charge retardation and interference.** (**A**) Schematic illustration of the simulation settings, where a longitudinal point dipole source is placed in a 2 nm gap layer between a 100 nm gold nanosphere and a flat gold substrate. (**B**) The Purcell factor spectrum of the point dipole source. (**C**) The electric field amplitude profile at 652 nm (the peak of the mode in the center of (B)), with the colors showing the value of $|E|$. (**D**) $E_z$ phase profile at 652 nm, with the colors showing the value of $\mathrm{Arg}(E_z)$. (**E**) Schematic illustration of the nanosphere surface charge density and phase profiles, for the linear scattering situation (left) and the nonlinear gap plasmon radiation situation (right). c is an arbitrary phase offset.